\documentclass[thmsa,final,a4paper,11pt]{article}
\usepackage{amssymb}
\usepackage{graphicx}
\usepackage{pppbib}
\begin{document}

\title{Gamma-Ray Bursts and Dark Matter \\
\noindent - a joint origin ?}
\author{Daniel Enstr\"{o}m\thanks{%
Email: daniele@mt.luth.se} \\
{\small Department of Physics }\\
{\small \ \ Lule\aa\ University of Technology}\\
{\small \ \ SE-971 87 Lule\aa , Sweden}}
\date{}
\maketitle

\begin{center}
\textbf{Contents:}
\end{center}

\hspace{2cm}1. Introduction

\smallskip

\hspace{2cm}2. The model

\smallskip

\hspace{2cm}3. Particle physics constraints

\smallskip

\hspace{2cm}4. Astrophysical constraints

\smallskip

\hspace{2cm}5. Conclusions and outlook

\section{\noindent Introduction}

Two of the most fascinating problems in astrophysics and cosmology today are
the origin and nature of the dark matter (DM) and the sources and mechanisms
behind the so-called gamma-ray bursts (GRB). The presence of non-luminous,
gravitationally interacting matter was inferred in the earlier part of this
century when the dynamics of groups of galaxies did not agree with the
predictions based on the observed amount of luminous matter. Ever since,
physicists have tried to build models that can explain both the nature and
the origin of the dark matter, in terms of either hot dark matter (HDM) or
cold dark matter (CDM), referring to their internal velocity distribution.
Recent analysis \cite{gawiser98} seems to favour a mixture of these two
categories, at least when compared to the observed cosmic microwave
background radiation (CMBR) and the standard theory for the large scale
structure formation in the Universe (arising from gravitational enhancement
of small density perturbations in the early Universe).

GRBs were first observed by the VELA satellites in the late 1960s \cite
{klebesadel73}. These bursts, isotropically distributed across the sky, are
assumed to originate from cosmological distances, and they radiate up to $%
10^{54}$ ergs in a time duration of $0.01-1000$ s. They are the most
luminous sources in the Universe. Several models have been proposed for the
origin of the bursts, and as of today, none has been singled out as most
favoured. So-called fireball models (see \cite{piran98} for an overview)
where a relativistic expanding shell, originating from the source, interacts
with the intergalactic matter is very popular since they can explain the
characteristics of the radiation. However, the actual source, the ``inner
engine'', of the energy release is not well understood. The most common
explanation includes a merger of two compact objects of some kind (neutron
stars, black holes etc.).

Our work is built on one, basic principle: to use known physics to explain
new phenomena. Our starting point has been the widely accepted theory that
in the early Universe, $t\lesssim 10^{-5}$ s after the Big Bang, all matter
was in the form of quark-gluon plasma (QGP). At $t\approx 10^{-5}$ s, a
phase transition occurred, which confined the free quarks into hadrons. This
quark-hadron transition is assumed to be of first order, something that
several lattice calculations seem to indicate \cite{boyd95}. In 1984 Witten 
\cite{witten84} suggested the possibility that regions of the high
temperature phase QGP would remain and be stable after the transition. If
this is correct, then a very exciting possibility to explain both the origin
of the baryonic dark matter and the sources for GRBs opens up. In our model,
we identify the sources of GRBs with the baryonic dark matter.

The fundamental assumption that underlies our model is that the QGP is the
absolute ground state of QCD. This idea is not new, in fact it emerged in
the early 1970s. An admixture of $u,\,d$ and $s$ quarks is likely to have a
lower energy per baryon number than the proton. This makes it possible for
the QGP to be stable, even on cosmological scales.

To summarize, we make two major assumptions in our model:

\begin{enumerate}
\item  QGP is the ground state of QCD, at least for massive objects.

\item  The quark-hadron transition was of first order.
\end{enumerate}

\section{The model}

The starting point for our model is the assumption that only a very small
part of the primordial QGP actually hadronised into ordinary matter. The
phase transition allowed QGP objects survive the transition, or ``quark
nuggets'' as other authors call them, fractally distributed in size and
occurring as inhomogeneities in the surrounding mixture of hadronic matter
and vacuum. Smaller QGP objects, with baryon number A $<10^{44}$ disappeared
due to neutron and proton condensation \cite{alam93}, made possible by the
high temperature, while larger objects survived.

Since the average density of one of these quark objects should be around $%
10^{13}$ g/cm$^{3}$, and the radius of an object with $10^{44}$ quarks is on
the order of a centimetre, gravity becomes important when deciding under
what circumstances the QGP objects are stable with respect to gravity and to
the internal degeneracy pressure occurring due to the Pauli principle. The
mass-radius relationship of a large spherical QGP bag (A $>10^{50}$) can be
calculated with the Tolman-Oppenheimer-Volkoff equations \cite{tov39},
derived in general relativity: 
\begin{equation}
\frac{dp}{dr}=-\frac{[\epsilon (r)+p(r)][m(r)+4\pi r^{3}p(r)]}{r[r-2m(r)]}
\label{eq1-TOV}
\end{equation}

\begin{equation}
\frac{dm(r)}{dr}=4\pi r^{2}\epsilon (r)\hspace{1cm}m(r)=4\pi
\int_{0}^{r}\epsilon (r^{\prime })r^{\prime }{}^{2}dr^{\prime }
\label{eq2-TOVmass}
\end{equation}

\begin{equation}
p(r=0)=p_{c}
\end{equation}

\begin{equation}
p(r=R)=0.
\end{equation}

Solving these equations (with $c=G=1)$ requires an equation of state for the
QGP. Our result is based on treating the QGP as a relativistic fermi gas
with zero chemical potential and no interactions. The phase boundary is
expressed in the spirit of the MIT bag model with an external bag pressure
characterised by the bag constant $B$: 
\begin{equation}
\epsilon (r)=3p(r)+4B.
\end{equation}
The kinetic part of the pressure is (with three quark flavours $f)$: 
\begin{equation}
p_{k}=\frac{8\pi ^{2}}{45}T^{4}+\sum_{f}\left( \frac{7}{60}\pi ^{2}T^{4}+%
\frac{1}{2}T^{2}\mu _{f}^{2}+\frac{1}{4\pi ^{2}}\mu _{f}^{4}\right) .
\end{equation}
When integrating the differential equation (\ref{eq1-TOV}), a mass-radius
relationship emerges, as shown in Figure 1.
\begin{figure}[h]
\centering
\includegraphics{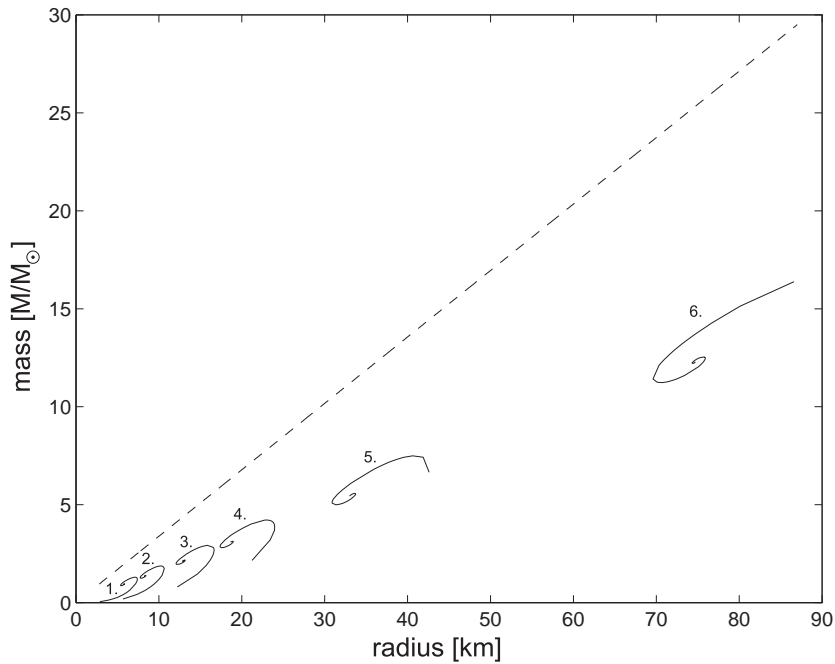}
\caption{\emph{The stability relations
(full lines) between the mass and the radius of an spherical QGP object for
different values of the external pressure }$B$\emph{. }$B^{1/4}=180$\emph{\
MeV (curve 1.), 150 MeV (2.), 120 MeV (3.), 100 MeV (4.), 75 MeV (5.) and 50
MeV (6.). For clarity, only the most relevant segments of the full lines are
shown. The hatched line shows the criterion for collapse into a black hole,
as given by the Schwarzschild radius. Results for other parameter values can
be found in \protect\cite{enstrom97}.}}
\label{massradius}
\end{figure}
Using these calculations for deciding the upper mass limit, QGP
objects surviving the quark-hadron transition and hadron evaporation must lie in the range $%
10^{-17}$ M$_{\odot }\,<$ m$_{QGP}<2$ M$_{\odot }$ with radius $%
10^{-5}$ km $<\,$r$_{QGP}<10$ km, or $10^{44}<\,$A$\,<10^{58}$. M$_{\odot }$
is the solar mass 2$\cdot 10^{33}$ g.

The size of the surviving QGP objects is characterised by the nucleation
distance in the phase transition. Ref. \cite{madsen96} suggests a nucleation
distance on the order of a cm, while others \cite{ignatius94} allow
distances on the order of a km. The assumption that the transition occurred
through detonation mechanisms \cite{abney94} has not been disproved, making
it impossible to exclude km-sized QGP remnants. The horizon size at the
transition is on the order of km with an enclosed mass of roughly M$_{\odot
} $. It is an interesting fact that the mass energy enclosed in a solar-mass
QGP object is roughly the same as the gamma-ray energy in a GRB at
cosmological distances, $z=O(1)$.

If the QGP objects survived the phase transition and the evaporation they
should be stable for a cosmological period of time, making up for at least
the baryonic part of the cold dark matter. The only way they could hadronise
or decay is if an external energy is added. One such circumstance is if two
of these objects merge and add gravitational energy to the system. This is
analogous to how a supernova can make elements heavier than iron. When a QGP
hadronise different radiative mechanisms \cite{harris96} might contribute
and gamma-ray emission is one of them. This is why we suggest that gamma-ray
bursts come about when parts of the QGP objects hadronise. One should notice
though that the mean free path for a gamma-ray in a quark-matter environment
is roughly 100 fm, which poses some difficulties in explaining how such a
vast amount of energy, 10$^{54}$ ergs, is radiated during a merger. This is
under further investigation, but one should notice that this problem is
shared by all proposed mechanisms for mergers as sources for GRBs. Our
scenario includes all the possible mechanisms for gamma-ray production
stated in the conventional scenarios and, \textit{in addition,} includes the
possibility of direct gamma-ray radiation through hadronisation.

One crucial point is of course the number density of QGP objects that merge.
We know from observations that the average galaxy hosts approximately $%
10^{-4}-10^{-6}$ bursts/yr and that must be made to fit our initial
distribution of solar mass sized QGP objects. An important fact is also that
all bursts seem to originate at cosmological distances. These two
observational facts suggest that mergers were more frequent in the distant
past and that the fraction of QGP objects (or at least of binary systems)
that lies in the M$_{\odot }$ range is fairly small.

We can conclude that there are no observational data that contradict the
idea of QGP objects as the sources for GRBs. Another consequence is that of
``beaming'' of the gamma rays, since the hadronisation in a merger probably
occurs through a bridge between the two objects. This also lowers the
required energy alleviating the problem mentioned above. This bridge forms
when the perturbing gravitational potential from the partner in the merger
is large enough to make it possible for the QGP to hadronise. This means
that the largest perturbation occurs along the symmetry line between the
centres of the two objects.

\section{Particle physics constraints}

One of our main assumptions is that the QGP is the true ground state of QCD.
This means that the energy content per quark is lower in a QGP than in a
nucleon. Such ideas began to flourish in the early 1970s with a pioneering
work by Bodmer \cite{bodmer71}. Since then several calculations supporting
this assumption have been made. Farhi and Jaffe \cite{farhi84} explored the
properties of the so-called strange matter and showed that within certain
parameter ranges, the energy per baryon in strange matter is lower than in
the proton. Witten \cite{witten84} showed that the presence of a $s$ quark
in a $u,\,d$ quark mixture lowers the energy per quark by a factor of 0.89
as compared to non-strange matter. Other analyses of the stability of
strange matter were performed by Bjorken and McLerran \cite{mclerran79},
Chin and Kerman \cite{chin79} and De R\'{u}jula and Glashow \cite{glashow84}.%
\begin{figure}[h]
\centering
\includegraphics[scale=0.5]{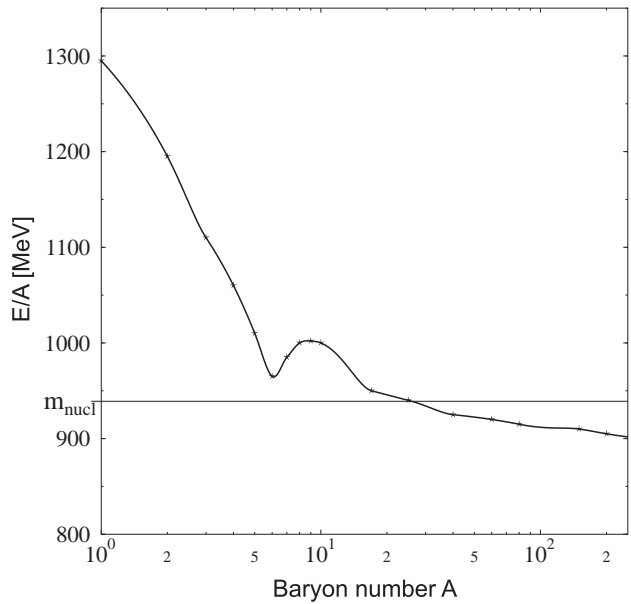}
\caption{\emph{This plot shows the
dependence of the energy ``per baryon'' of a giant bag on the baryon number
for }$B^{\frac{1}{4}}=145$\emph{\ MeV \protect\cite{greiner98}. The matter
in the bag is SU(3) flavour symmetric and electrically neutral.}}
\label{fig2E/A}
\end{figure}
They suggested that since the mass of the $s$ quark is lower than the Fermi
energy of a $u$ or $d$ quark in a multiquark system, the opening of a new
flavour degree of freedom could lower the Fermi energy of the system. This
behaviour can be seen in the simulation shown in Figure 2. Other simulations
have shown similar results but no such stable multiquark objects have been
detected. The existence of such a quark matter phase is hence just a
theoretical prediction. The reason why atomic nuclei do not decay into a QGP
is that the conversion of a $u$ or a $d$ quark into an $s$ quark is a weak
process with a very low probability. The transition of $^{56}$Fe into
strange matter containing 168 $u$, $d$ and $s$ quarks in equal proportions
is a 56th order weak process. Hence, the probability of decay of ordinary
matter into strange matter is exceedingly small. It should be noted that it
is not obvious that models of confinement, such as the MIT\ bag model \cite
{chodos74}, can be used for bags of arbitrary sizes or that the bag constant
must be independent of the number of confined quarks.

The details of the quark-hadron transition are not very well known. Large
experiments (either running or planned) will try to detect the phase
transition. The consensus among both cosmologists and particle physicists is
that it occurs at a temperature of $\sim 150$ MeV and is of first order.
Lattice calculations seem to indicate a first order transition at that
temperature but it is fair to say that the results are inconclusive, due to
tremendous difficulties of simulating a fermion system with a non-zero
chemical potential. This applies both to the type of the transition, the
temperature and the mean nucleation distance.

The possible survival of QGP objects has been discussed for some time. Some
authors have argued \cite{alcock89} that large QGP objects would boil when
hadronic gas is formed in their interior due to the superheating they are
experiencing. However, the importance of this mechanism has been questioned 
\cite{madsen91}, since the nucleation of hadronic bubbles in the interior is
a rather slow process compared to surface evaporation and cooling.

\section{Astrophysical constraints}

Two different observations will be discussed here. One of these is the CMBR
which comes from the last scattering surface, $\sim 10^{5}$ years after the
Big Bang. During the last nine years, the COBE satellite has measured the
CMBR all over the sky. The data describe an almost isotropic, black-body
radiation at a temperature of T $=2.725\pm 0.005\,$K \cite{mather90}. The
small anisotropies in the radiation distribution, $\frac{\delta T}{T}\sim
30\pm 5\,\mu $K, is assumed to originate from small density perturbations
present at the time of the last scattering, since the anisotropy is related
to fluctuations in the density distribution, 
\begin{equation}
\frac{\delta T}{T}\propto \frac{\delta \rho }{\rho }.
\end{equation}
If our scenario of surviving large QGP objects is correct, then one might
think that this is not compatible with the COBE data due to the huge
``lumpiness''. This is not the case since only non-neutral matter interacts
with the photons present at the time of last scattering. Since the QGP
objects are charge-neutral made up of $u$, $d$ and $s$ quarks and a small
amount of electrons, the possible imprint left on the CMBR by the QGP
objects is most likely unobservable. We would like to add, however, that
this is not a settled issue and a thorough investigation of the exact
interaction between the CMBR and the QGP objects remains to be made.

The other major observation we want to address is the abundances of light
elements. The standard theory (``SBBN'') describing the formation of these
elements states that they appeared when the Universe was between $10^{-4}$ s
and a few minutes old. Since the QGP objects in our model are in their
ground state, these did not partake in the formation of light elements. The
presence of large inhomogeneities in the nucleosynthesis era affects the
nucleosynthesis. Various authors \cite
{applegate87,alcock87,malaney88,kurki88} have addressed inhomogeneous Big
Bang nucleosynthesis and recent results \cite{kurki98} indicate that the
presence of inhomogeneities on the metre to kilometre scale actually reduces
the discrepancy between the observed abundances of $^{4}$He and D and the
prediction of SBBN. Ref. \cite{kurki98} states that the origin of these
inhomogeneities could very well be the quark-hadron transition.

\section{Conclusions and outlook}

One can conclude that no observations rule out the possible presence of
relic QGP objects, acting as both the baryonic cold dark matter and as the
source for gamma-ray bursts. However, one should also say that no direct
observations support this idea. The two assumptions we have made regarding
the stability of QGP and the nature of the phase transition have no
experimental evidence either. There is no doubt that our model is to be
considered as both rudimentary and speculative. Much work remains to be
done, where perhaps the most important one is pure particle physics; to
establish whether or not the QGP is the ground state of QCD. The results of
future experiments at RHIC and LHC will hopefully confirm the existence of a
deconfined phase, the QGP. On the theoretical side, a detailed analysis of
the split-up of the QGP phase in the quark-hadron transition is desireble.

On the observational side there is certainly an enormous amount of
information in the characteristics of the CMBR remaining to be discovered,
which improved data taking and analysis should be able to extract. For this,
a more detailed analysis of the possible interaction between the QGP objects
and the CMBR is necessary.

This work has been done in collaboration with Sverker Fredriksson and Johan
Hansson at the Department of Physics, Lule\aa\ University of Technology,
Svante Ekelin at the Department of Mathematics, Royal Institute of
Technology, Stockholm and Argyris Nicolaidis at the Department of
Theoretical Physics, Aristotle University of Thessaloniki.

\textbf{Acknowledgments:} The author would like to thank the Organisers, and
in particular Professor Zichichi, for hospitality and for arranging a most
rewarding school. He is also grateful to Professors Witten, Roos and Guidice
for very constructive comments.

\end{document}